\documentclass[preprint]{elsarticle}

\usepackage{amssymb}

\journal{arXiv}
\begin{document}

\begin{frontmatter}



\title{Observation of isomeric decays and the high spin states in doubly-odd $^{208}$Fr}

\author[sinp]{D.~Kanjilal}

\author[sinp,capp]{S.~Bhattacharya}
\author[sinp]{A.~Goswami}
\author[sinp]{R.~Kshetri}
\author[sinp]{R.~Raut}
\author[sinp,capp]{S. Saha\corref{cor1}}
\ead{satyajit.saha@saha.ac.in}
\author[iuac]{R.~K.~Bhowmik}
\author[iuac]{J.~Gehlot}
\author[iuac]{S.~Muralithar} 
\author[iuac]{R.~P.~Singh}
\author[andhra]{G.~Jnaneswari}
\author[vec]{G.~Mukherjee}
\author[visv]{B.~Mukherjee}

\cortext[cor1]{Corresponding author}

\address[sinp]{Nuclear and Atomic Physics Division, Saha Institute of Nuclear Physics, Kolkata 700064, India}

\address[capp]{Centre for Astro-Particle Physics, Saha Institute of Nuclear Physics, Kolkata 700064, India}

\address[iuac]{Inter University Accelerator Centre, New Delhi 110067, India}

\address[andhra]{Department of Physics, Andhra University, Vishakhapatnam 530003, India}

\address[vec]{Variable Energy Cyclotron Centre, Kolkata 700064, India}

\address[visv]{Department of Physics, Visva Bharati, Santiniketan 731235, India}

\begin{abstract}
Neutron deficient isotopes of Francium ($Z$=87, $N \sim 121-123$) as excited nuclei were 
produced in the fusion-evaporation reaction: $^{197}$Au~($^{16}$O, $xn$)~$^{213-x}$Fr 
at 100 MeV. The $\gamma$ rays from the residues were observed through the high 
sensitivity Germanium Clover detector array INGA. The decay of the high spin states 
and the isomeric states of the doubly-odd $^{208}$Fr nuclei, identified from the known 
sequence of ground state transitions, were observed. The half lives of the 
$E_\gamma =$~194(2) keV isomeric transition, known from earlier observations, was 
measured to be $T_{1/2}=$~233(18) ns. A second isomeric transition at 
$E_\gamma =$ 383(2) keV and $T_{1/2} =$~33(7) ns was also found. The measured half 
lives were compared with the corresponding single particle estimates, based on a 
the level scheme obtained from the experiment.  

\end{abstract}

\begin{keyword}

\PACS 21.10.Tg \sep 23.20.Lv \sep 23.35.+g \sep 27.80.w

\end{keyword}
\end{frontmatter}


\section{Introduction}
\label{sc:intr}
Investigation on the nuclear structure of the trans-Lead neutron deficient nuclei ($Z > 82$, 
$N<126$) have attracted much attention in recent years\cite{hart,may,pod}. For many of 
these nuclei, only the ground state spin and parity are known from their $g$-factor/magnetic 
moment measurements, and perhaps a few low lying excited states are observed so far. 
The major difficulties in populating the high spin states in these nuclei are: a) very low 
cross sections for the formation of evaporation residues~(ER), and b) very high probability of 
fission, which removes flux from the ER channel and prevents the excited nucleus from 
sustaining large angular momenta needed to populate the high spin states.

Experimental investigation of the high spin states of quite a few trans-Lead neutron 
deficient nuclei have been of interest recently. A series of investigations on 
the $^{211-214}$Fr isotopes\cite{byrne1,byrne2} showed that the structure can be 
interpreted in terms of the shell model states, and the excited states reveal an 
interplay between the protons in the $(1h_{9/2}, 2f_{7/2}, 1i_{13/2})$ states and 
the neutron holes in the $(2f_{5/2},3p_{3/2},1i_{13/2})$ orbitals, or the neutrons 
promoted to the $(2g_{9/2},1i_{11/2},1j_{15/2})$ high spin orbitals by core excitation, 
leading to the generation of high spin states. One of the major interests in the 
spectroscopic investigation of these nuclei is the role played by the 
$i_{13/2}$ shell in creating isomeric levels which decay through transitions of 
higher multipolarity or are hindered by the close proximity of the levels below.

A few spectroscopic investigations on the proton rich lighter Francium isotopes have 
been made recently. While a complete study of the $^{205-207}$Fr nuclei\cite{hart}, 
using Gammasphere and the HERCULES II array for filtering out the evaporation residues 
from the fission background, revealed the existence of shears band in $^{207}$Fr~($N=120$), 
investigations on the $^{208-210}$Fr nuclei\cite{may,pod} have resulted in contradictory 
conclusions. Spectroscopic studies made using the YRAST BALL array, comprising six 
Compton suppressed Clover HPGe detectors\cite{may}, coupled to the SASSYER recoil 
separator\cite{ress} for selecting the evaporation residues had concluded that the 
pair of intense gamma rays of 632 keV (ground state transition) and 194 keV (isomeric
transition) belong to $^{209}$Fr. The half life of the isomeric transition was 
measured to be 446(14)~ns. At the same time, another independent study of isomeric 
decay of proton rich nuclei produced by projectile fragmentation reaction of $^{238}$U 
beam at 900 MeV/u on $^{9}$Be target at the Fragment Recoil Separator (FRS) facility 
of GSI, Darmstadt, Germany had assigned the same pair of gamma rays to 
$^{208}$Fr\cite{pod}. Clean isotopic and isobaric resolution of nuclei, which have 
low lying isomers with half lives $\gtrsim$ 100 ns, are routinely achieved using this 
facility\cite{gei}. Half life of the 194 keV isomeric transition was reported to 
be $\sim 200$~ns, though the prompt transitions above the isomer could not be observed 
because of experimental constraints. This paper reports investigation of the 
isomeric transitions in $^{208}$Fr, along with the high spin states above the 
isomers. It is noteworthy to report here that at the time of final phase of data 
analysis and preparation of the manuscript, a paper on assignments of levels in 
$^{208}$Fr was published by G. D. Dracoulis {et al.}\cite{drac}. The 
differences in our methods, the results of observations on the isomeric 
transitions, new $\gamma$ transitions over and above those reported therein, 
and the basis for our assignments of the isomeric transitions 
to $^{208}$Fr are highlighted in this paper.

\section{Experiment and data analysis}
\label{sc:expt}
The experiment to produce $^{208}$Fr  was carried out at the 
Inter-University Accelerator Centre (IUAC), New Delhi. 
The Fr isotopes were produced by bombarding a 3.5 mg.cm$^{-2}$ self-supporting 
Gold (99.95\% purity) target with $^{16}$O beam at 88, 94 and 100 MeV. The nuclei 
of interest were produced as evaporation residues (ER) through ($^{16}$O,~$xn\gamma$) 
reactions. The target thickness was chosen on the basis of energy loss calculations 
based on SRIM2003\cite{srim} to stop more than 90\% of the ER within the target, and 
allow most of the fission fragments to fly away from it. Estimation of cross sections, 
angular distributions of the evaporation residues (ER) and the fission yield were done 
using the code PACE\cite{pace}. Based on these calculations, $\sim 60-80$~\% of the 
fusion products at these bombarding energies undergo fission, which causes a huge 
background. Therefore, an effective filter to clean up the spectra, and / or good 
statistics are essential for extraction of meaningful results, as were done 
recently\cite{hart}. In the absence of such filters, use of large gamma detector arrays 
with high resolving power, good efficiencies for $\gamma \gamma$ and 
$\gamma \gamma \gamma$ coincidence events, along with additional measurements of 
excitation functions from in-beam and off-beam measurements yielding consistent 
results, have been utilized in our attempt to resolve the ambiguity. 
\begin{figure}
\begin{center}
\includegraphics[scale=1.0]{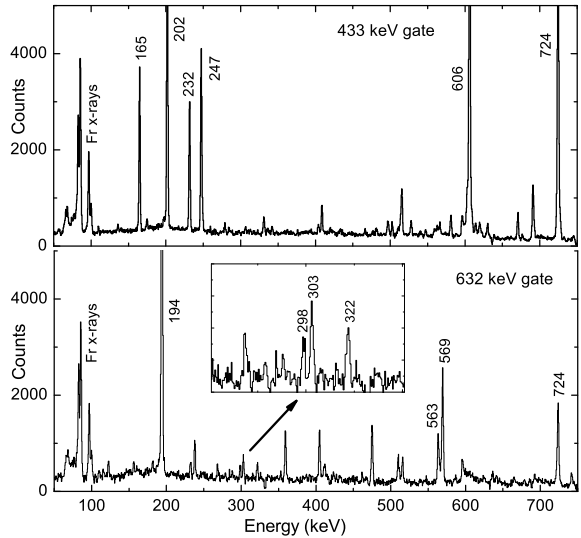} 
\end{center}
\caption{\label{fg1} A few gated spectra of the Fr isotopes.}
\end{figure}

The $\gamma$-rays produced were detected by the Indian National Gamma Array (INGA) 
consisting of 18 Compton suppressed Clover Germanium detectors placed around the 
target\cite{inga1} at the INGA-HYRA beam line of the IUAC, New Delhi. Four Clover 
detectors were placed at $57^\circ$, six at $90^\circ$, four at $123^\circ$, and 
four at $148^\circ$ for facilitating measurement of directional correlation 
from oriented states (DCO) ratios. The linear signals, along with the anti-coincidence 
logic and the trigger signals, were processed through the indigenously developed 
Clover electronics modules dedicated for the INGA set-up\cite{clov-elec}. Since the 
measurement of half lives of the isomeric states is crucial for the experiment, time to 
digital converters (TDC) were used with stop signals from the individual Clover units 
and common start signal from the master trigger which can be chosen to select 2-fold or 
higher fold events. Range of the TDC was set to 400 ns for the exclusion of delayed 
$\gamma$-rays possibly coming from $\alpha$- and $\beta$ decays. Altogether 
$315 \times 10^{6}$ two-fold and $48 \times 10^{6}$ three-fold coincidences were 
recorded in $\sim$50 hours at 100 MeV beam energy, and $\sim$~20\% of above numbers 
were obtained at 88 and 94 MeV. In order to identify and measure the yield of the ER 
nuclei from their known $\alpha-$ and $\beta-$decay modes, data were taken in the 
multiscaling mode during the beam-off condition between the runs. All the on-line 
and off-line data were collected using the CAMAC-based CANDLE data acquisition 
system\cite{candle}, and were analysed off-line using CANDLE, INGASORT\cite{inga2} 
and RADWARE\cite{rad} analysis softwares.
\begin{figure}
\center{
\includegraphics[scale=1.2]{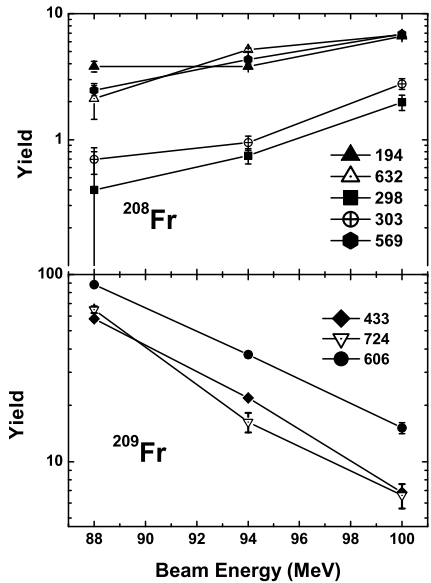} 
}
\caption{\label{fg2} Excitation functions measured by tagging on the strong 
gamma transitions.}
\end{figure}

Data analyses were done in several steps. For the excitation function measurements, 
$\gamma \gamma$ coincidence matrices were generated from the 2-fold data collected 
at the three different energies. Yields for the intense $\gamma$-rays, normalized 
by the current integrator readings recorded at the beam dump during the excitation 
function runs, were obtained from the Francium $K_\alpha$ and $K_\beta$ x-ray gated 
projections of the symmetrized matrices for each energy. The gated spectra are shown 
in the Fig.~\ref{fg1}, where the intense $\gamma$-rays of $^{208}$Fr and $^{209}$Fr 
are indicated. The 724 keV $\gamma$-ray is present in both the nuclei, and hence 
could not be resolved in the experiment. Since its intensity is stronger in the 
decay of $^{209}$Fr levels, the excitation function shows the energy dependence 
typical of $^{209}$Fr. The trends of the excitation functions graphs for $^{208}$Fr 
and $^{209}$Fr, shown in the Fig.~\ref{fg2} are comparable to that predicted from 
the PACE calculations, though it overestimates the $^{208}$Fr yield and underestimates
the $^{209}$Fr yield at 100 MeV. The $^{210}$Fr yield could not be measured even at 
the two lower energies as the corresponding characteristic $\gamma$-rays are not 
known.

\begin{figure}
\begin{center}
\includegraphics[scale=1.2]{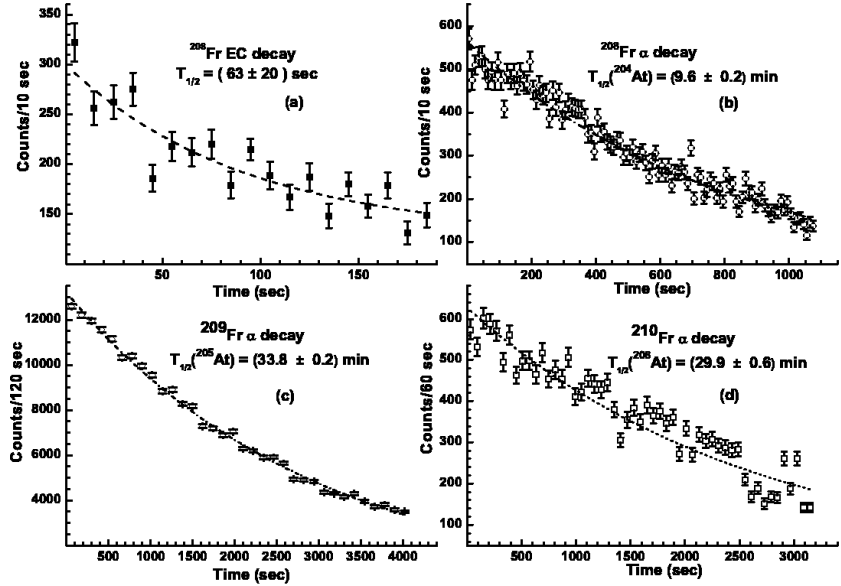} 
\end{center}
\caption{\label{fg3} Offline decay plots of various isotopes used for extracting the yield ratio
of the Francium evaporation residues.}
\end{figure}

The relative yields of the Fr isotopes ($A=208-210$) at the three different energies 
were also obtained from the radioactive decay runs taken in the beam-off condition. The time 
stamps in the list data blocks were used to generate the time marker. The $E_\gamma$ vs. time 
matrices were generated using the CANDLE analysis software from which, the $E_\gamma$ spectra 
at different time windows were generated. The decay curves for the intense characteristic 
$\gamma-$rays belonging to the decay branches of these nuclei were obtained. A few decay curves 
are shown in the Fig.~\ref{fg3}. The ERs decay via their well known $\alpha-$ and $\beta-$decay 
branches, with half lives which are well documented for these nuclei. The half lives $T_{1/2}$, 
obtained by fitting the equation: $n=n_0 \exp(-0.693\,t/\,T_{1/2})$ for different decay branches 
observed in this experiment, are enlisted in the Table~\ref{tb1}. The results show reasonably 
good agreement with the reference values quoted in the NNDC database\cite{mart1,sch1,mart2,kon1,brow,kon3}. 
Relative yields of different ERs at the beginning of beam-off runs, were extracted from the 
fitted $n_0$ values, measured half lives and the branching fractions of the different decay 
branches. The relative yields, obtained from our data for the Fr isotopes, are shown in the 
Table~\ref{tb1}. The major contribution to the uncertainties ($\lesssim 5\%$) quoted in the relative 
yields is from the fluctuation in the beam current before the beginning of the beam-off runs, 
which were monitored during the analysis, and the rest from the exponential fit to the data. 
The results are in reasonable agreement with the measured excitation functions of Fig.~\ref{fg2}, 
which agrees with the assignment of the 826 keV isomeric level and the corresponding gamma 
transitions to $^{208}$Fr, rather than to $^{209}$Fr\cite{may}, in agreement with the 
conclusions of Refs.~\cite{pod,drac}.  
\begin{table*}
\caption{Half lives and relative yields of Fr isotopes from the off-line decay runs. 
Percent figures within brackets in col.~2 are the branching fractions. The 
characteristic $\gamma$-transitions $E_\gamma$ (col.~5), monitored for decay analysis, are the 
strongest lines in the corresponding daughter nuclei. The yields given are the 
extracted relative yields at each beam energy.}
\begin{tabular}{cccccccc}
\hline\hline
ER	& Decay branch & $T_{1/2}$ & $T_{1/2}$ & $E_\gamma$ & Yield 
& Yield & Yield \\
& & & (This expt) & (keV) & (88 MeV) & (94 MeV) & (100 MeV) \\ \hline 
$^{208}$Fr & $^{208}$Fr $\rightarrow$ $^{208}$Rn (11\%) & 59.1(3) sec\cite{mart1} & 63(20) sec & 635.8 & & & \\
           & $^{208}$Fr $\rightarrow$ $^{204}$At (89\%) &             &          &       & & & \\
           & $^{204}$At $\rightarrow$ $^{204}$Po (96\%) & 9.2(2) min\cite{sch1}  & 9.6(2) min & 516.3 & 0 & 0.14(1) & 0.46(8) \\ \hline
$^{209}$Fr & $^{209}$Fr $\rightarrow$ $^{209}$Rn (11\%) & 50.0(3) sec\cite{mart2} & & & & & \\
           & $^{209}$Fr $\rightarrow$ $^{205}$At (89\%) &             &          &       & & & \\
           & $^{205}$At $\rightarrow$ $^{205}$Po (90\%) & 26.9(8) min\cite{kon1}  & 33.8(2) min & 719.3 & 1.0(2) & 1.00(5) & 1.0(3) \\ \hline
$^{210}$Fr & $^{210}$Fr $\rightarrow$ $^{210}$Rn (40\%) & 3.18(6) min\cite{brow} & 3.4(2) min & 643.8 & & & \\
           & $^{210}$Fr $\rightarrow$ $^{206}$At (60\%) &             &          &       & & & \\
           & $^{206}$At $\rightarrow$ $^{206}$Po (99\%) & 30.0(8) min\cite{kon3}  & 29.9(6) min & 700.7 & 0.35(5)& 0.18(1) & 0.37(7)\\
\hline \hline
\end{tabular}
\label{tb1}
\end{table*}

From the online data taken at 100 MeV beam energy, the $\gamma-\gamma$ matrices, 
Francium $X$-ray gated $\gamma \gamma$ matrices, the prompt and the delayed 
$\gamma \gamma$ matrices and the $\gamma$-gated $\gamma \Delta T$ matrices were 
constructed for establishing the level scheme and resolving the isomeric 
transitions. From the first two sets of matrices and by gating on the strongest 
632 keV ground state transition and the intense 194 keV transition, the 
$\gamma$-transitions belonging to $^{208}$Fr were clearly established. These 
transitions, observed at 100 MeV beam energy, are shown in the Table~\ref{tb2}. 
The relative intensities given in the table were obtained from the 
632 keV gated spectra. Because of the existence of low lying isomer with half lives
$\sim 200-400$ ns, and also due to the large internal conversion of some of the 
levels, the intensity balance across the isomeric levels could only 
approximately be done. 
\begin{table*}
\caption{Intensities and DCO ratios of the $\gamma$-rays of $^{208}$Fr at 100 MeV beam 
energy. Intensities are normalized relative to the 632 keV ground state 
transition. Newly found $\gamma$-transitions in this experiment are indicated by $^a$. 
For the DCO ratios, the gating transitions are indicated in col.~6. Multipolarity of 
the gating transitions are indicated in brackets and the multipolarity assignments are 
shown in col.~7.}
\begin{tabular}{lrrrrrl}
\hline \hline
$E_{\gamma}$ & $I_{\gamma}$ & $E_i$ & $E_f$ & $R_{\rm DCO}$ & $E_{\gamma}$ (gate) & Multipolarity \\
(keV) & &(keV) &(keV)& & (keV) &  \\ \hline
110.0(20)$^a$ & 11.00(45) & 4077 & 3967 & & &\\
115.8(20)$^a$ & 7.58(31)  & 3967 & 3851 &&&\\
122.7(18)$^a$ & 17.9(13)  & 3851 & 3728 &&&\\
123.2(14)$^a$ & 8.6(10)   & 2262 & 2139 &&&\\
150.2(17)$^a$ & 8.3(17)   & 3856 & 3706 &&&\\
156.4(17)$^a$ & 13.4(18)  & 2139 & 1983 &&&\\
160.1(17)     & 9.9(25)   & 4070 & 3910 & 1.26(59) & 303({\sl M1}) & {\sl M1} \\
181.8(18)$^a$ & 11.6(19)  & 3910 & 3728 & 1.20(39) & 303({\sl M1}) & {\sl M1} \\
194.1(18)     & 818(12)   & 826  & 632  & 0.99(6) & 632({\sl E2}) & {\sl E1} $(9^{-} \rightarrow 9^{+})$ \\
209.7(19)$^a$ & 8.70(79)  & 2763 & 2553 & $\sim 0.5$  & 303({\sl M1}) & {\sl E2} \\
232.4(18)     & 15.9(13)  & 4325 & 4093 & 1.93(80) & 569({\sl E2}) & {\sl M1} \\
238.2(18)     & 38.8(16)  & 4093 & 3855 & 1.74(32) & 569({\sl E2}) & {\sl M1} \\
268.5(14)$^a$ & 15.4(20)  & 1983 & 1715 & & &\\
284.6(15)$^a$ & 5.0(16)   & 3048 & 2763 & $\sim 0.5$ & 303({\sl M1}) & {\sl E2} \\
288.1(18)     & 11.39(4)  & 2246 & 1958 & $\sim 0.5$ & 569({\sl E2}) & {\sl M1} \\
290.9(15)$^a$ & 3.8(13)   & 2553 & 2262 & $\sim 0.5$ & 303({\sl M1}) & {\sl E2} \\
298.8(19)     & 17.1(28)  & 2133 & 1834 & 1.04(13) & 303({\sl M1}) & {\sl M1} \\
303.1(19)     & 26.8(35)  & 1512 & 1209 & 2.1(6) & 632({\sl E2}) & {\sl M1} \\
313.1(17)$^a$ & 3.9(12)   & 3728 & 3415 & $\sim 0.5$ & 303({\sl M1}) & ({\sl E2}) \\ 
321.8(22)     & 24.8(12)  & 1834 & 1512 & 1.02(18) & 303({\sl M1}) & {\sl M1} \\
359.3(21)     & 67.6(41)  & 1715 & 1356 & 1.03(21) & 632({\sl E2}) & {\sl E2}\\
365.4(16)$^a$ & 4.5(12)   & 2633 & 2268 & & &\\
382.9(18)$^a$ & 7.13(90)  & 1209 & 826  & 0.60(17) & 322({\sl M1}) & {\sl E2} \\
404.8(19)     & 64.9(39)  & 1800 & 1395 & 0.92(12) & 569({\sl E2}) & {\sl M1/E1} $(11 \rightarrow 11)$ \\
410.7(13)$^a$ & 8.45(76)  & 3415 & 3004 & $\sim 0.5$ & 303({\sl M1}) & ({\sl E2}) \\
428.3(15)$^a$ & 7.0(13)   & 2262 & 1834 & 0.50(34) & 303({\sl M1}) & {\sl E2} \\
467.8(21)$^a$ & 9.61(41)  & 2268 & 1800 & 0.8(4)& 405({\sl M1/E1}) & {\sl M1/E1} $(11 \rightarrow 12)$ \\
498.6(13)$^a$ & 4.5(12)   & 3706 & 3207 & & &\\
510.2(22)     & 40.6(33)  & 3190 & 2680 & $\sim 2$ & 569({\sl E2}) & {\sl M1} \\
563.5(24)     & 88.4(45)  & 1958 & 1395 & 1.85(44) & 632({\sl E2}) & {\sl M1} \\
566.2(24)$^a$ & 27.1(38)  & 5487 & 4921 & 0.44(17) & 563({\sl M1}) & {\sl E2} \\
569.4(24)     & 197.7(62) & 1395 & 826  & 0.89(9) & 632({\sl E2}) & {\sl E2} \\
596.2(24)$^a$ & 38.5(47)  & 4921 & 4325 & 2.0(8) & 632({\sl E2}) & {\sl M1} \\
603.6(26)$^a$ & 12.3(26)  & 3207 & 2603 & & &\\
632.2(19)     & 1000      & 632  & 0    & & & {\sl E2} \\
645.3(19)$^a$ & 8.9(31)   & 2603 & 1958 & 0.6(4) & 563 ({\sl M1}) & ({\sl E1/M1}) ($12 \rightarrow 12$) \\
665.1(20)$^a$ & 11.7(47)  & 3855 & 3190 & 1.05(37)& 563({\sl M1}) & {\sl M1} \\
721.6(23)$^a$ & 21.0(38)  & 2680 & 1958 & 1.0(6) & 563({\sl M1}) & {\sl M1} \\
723.9(23)     & 160.30(46)& 1356 & 632  & 1.14(11)& 632({\sl E2}) & {\sl E2} \\
742.0(23)     & 32.2(13)  & 3004 & 2262 & 0.56(23) & 303({\sl M1}) & {\sl E2} \\
774.6(18)     & 5.28(2)   & 2169 & 1395 & $\sim 1$ & 569({\sl E2}) & ({\sl E2}) \\
880.2(16)     & 7.26(2)   & 2680 & 1800 & $\sim 1$ & 569({\sl E2}) & ({\sl E2}) \\
955.8(16)$^a$ & 3.6(11)   & 4004 & 3048 & 0.49(24) & 303({\sl M1}) & ({\sl E2}) \\ 
\hline \hline
\end{tabular}
\label{tb2}
\end{table*}
The observed $\gamma$-rays and their relative intensities match reasonably well with those obtained 
recently by Dracoulis {et al.}\cite{drac}. However, quite a few additional $\gamma$-transitions 
were observed and indicated in the table. A directional correlation of oriented nuclei (DCO) analysis
was also performed with the data taken at $(90^\circ, 148^\circ)$ and $(90^\circ, 123^\circ)$ angle pairs. 
The gating transitions and their multipolarities used for DCO ratio calculations are shown in the 
Table~\ref{tb2}. These assignments match with those obtained in Ref.~\cite{drac}.

\begin{figure}
\begin{center}
\includegraphics[scale=1.0]{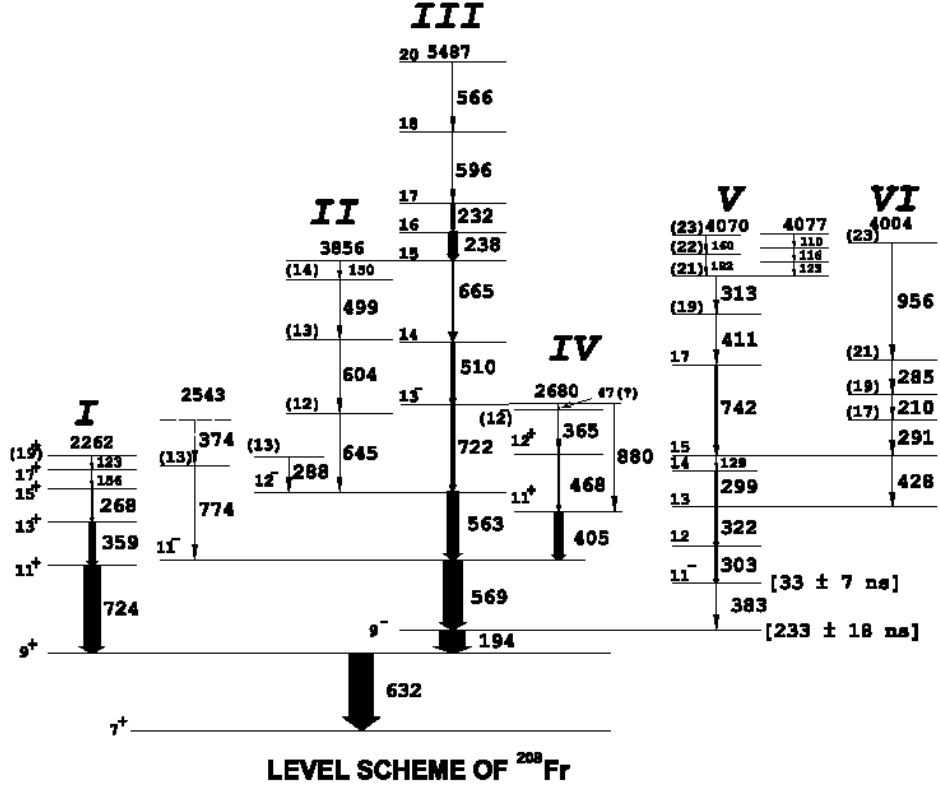} 
\caption{\label{fg4} Level scheme of $^{208}$Fr obtained from the present work. Transitions which were 
not observed but necessary to fit the level scheme are indicated by ?. The dotted transitions are 
observed in the previous work\cite{drac}, but not in this work.}
\end{center}
\end{figure}

\begin{figure}
\begin{center}
\includegraphics[scale=1.2]{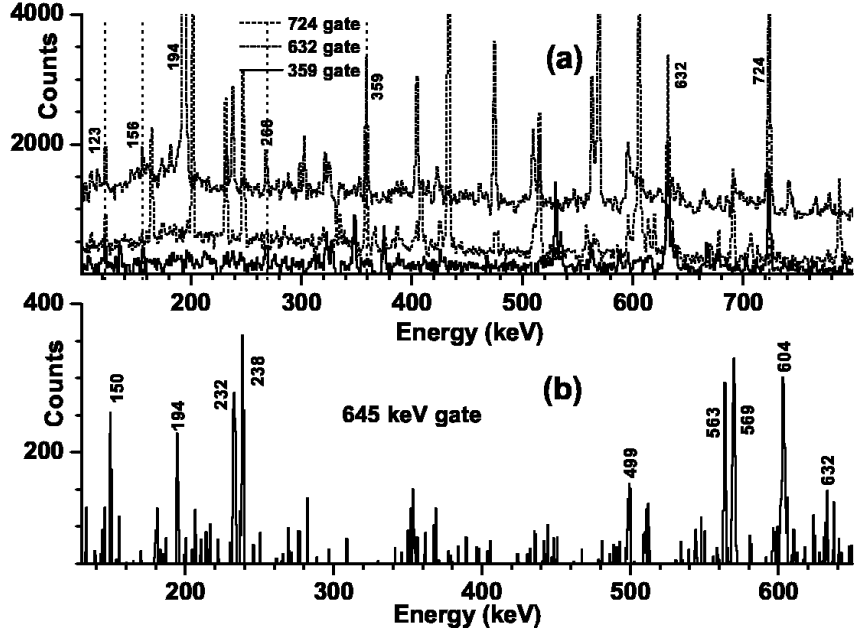} 
\caption{\label{fg5} (a) Overlap of the 359, 632 and 724 keV gated spectra manifesting the sequence I
transitions above the 1750 keV level. (b) The 645 keV gated spectrum for the sequence II transitions.}
\end{center}
\end{figure}

\begin{figure}
\begin{center}
\includegraphics[scale=1.3]{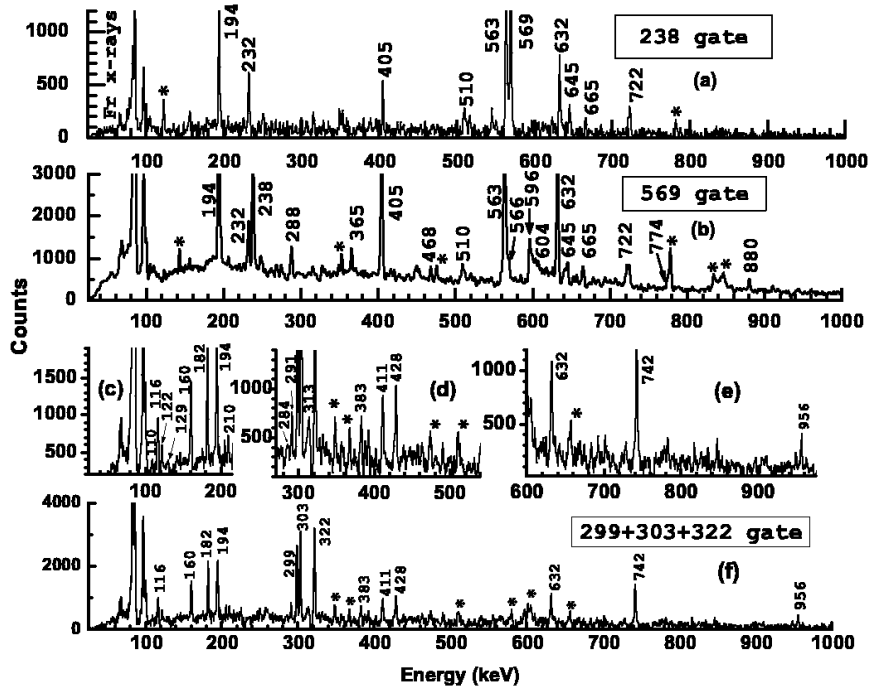} 
\caption{\label{fg6} A few relevant gated spectra of $^{208}$Fr. Gating transitions are 
indicated in the figures. Contamination lines are indicated by $*$.Plots in (c) to (e) 
are zoomed plots of the spectra shown in (f), to bring out the weak transitions noted 
in the experiment.}
\end{center}
\end{figure}

Based on the intensity correlations obtained from our gated spectra, and also from the DCO ratio 
measurements, the level scheme for $^{208}$Fr was established as shown in the Figure~\ref{fg4}. 
Apart from one sequence of transitions (I) passing through 359 and 724 keV, which directly feeds the 
632 keV first excited state, two major sequences (III and V) and three minor sequences 
(II, IV and VI) of transitions, which pass through the isomeric 826 keV level, have been 
observed. A few relevant gated spectra are shown in the Figs.~\ref{fg5} and \ref{fg6}. 
Out of these, three sequences (II, III and IV) of transitions pass 
through the strong 569 keV transition, and two (V and VI) through the sequence of 303 and 
322 keV transitions. In all the major sequences, ordering of the transitions are cross-checked 
by intensity correlations and also by reverse gating. 

About 25 new transitions, over and above those observed by Dracoulis et al.\cite{drac}, were found, 
as noted in the Table~\ref{tb2}. The 722 keV transition in sequence III falls on the tail of the 
much stronger 724 keV transition of sequence I, and could only be observed in the 238 and 569 keV 
gates, as shown in the Fig.~\ref{fg6}, except for a small contamination at 725 keV in the 569 keV 
gate, possibly coming from $^{198}$Au produced by neutron transfer reaction. The 725 keV 
contamination was also present in the 194 keV gated spectrum for the same reason mentioned 
above. The 724 keV contamination line was not found in the 563 keV gate. The contamination from 
the 724 keV transition is much more severe as it also belongs to $^{209}$Fr. Similarly, the 
510 keV $\gamma$-ray is also a new line, which was observed consistently in the coincident 
gates of sequence III transitions. It was placed in the level scheme just above the 722 keV 
transition, based on the observed intensities in various forward and reverse gates. The 
665-510-722 keV sequence in III is bypassed by the 150-499-604-645 keV sequence II transitions 
which is evident from the 645 keV gated spectrum of Fig.~\ref{fg5}~b. The 665 keV and the 
150 keV transitions are weak in intensity and allow only an approximate intensity balance. 
The sequence I was known up to 359 keV\cite{drac}, which was extended with the 123-156-268 keV 
transitions observed by overlapping the 632,724 and 359 keV gated spectra (see Fig.~\ref{fg5}a). 
The sequence IV transitions pass through 569 keV but they bypass the 563 keV and 722 keV 
transitions and hence were placed accordingly. The 880 keV transition, which was also shown in 
Ref.~\cite{drac}, was absent in 365 and 468 keV gates, but present in the 405 keV gate. Hence 
it was placed to directly feed the 1800 keV level. By matching the sum energies of the relevant 
transitions in IV, a 47 keV transition was placed just below the 2680 keV level. This, perhaps, 
could not be identified due to the absorbers placed in front of the Clover detectors. 

For the sequence V and VI, several new transitions were found and the relative ordering was 
modified over the level scheme shown in Ref.~\cite{drac}. 210, 284, 291 and 956 keV lines were 
observed for the first time in the 299, 303 and 322 keV gates (see Fig.~\ref{fg6}a), but absent 
from the 742 keV gate. This makes the placement of 742 keV above the 299-303-322 keV sequence. 
Further, the 428 keV line was present in the 303, 322 and 742 keV gates but absent in the 299 
keV gate. This was also confirmed by reverse gating on 428 keV. The 129 keV line, though weak 
possibly because of internal conversion, was present in 299, 303 and 322 keV gates but absent 
in the 428 keV gate, which makes its placement sequential to 299 keV but parallel to 428 keV. 
The $\gamma$-rays coming from the levels above 3004 keV in sequence V were observed for the 
first time, except for the 160 keV transition which has already been reported\cite{drac}. 
A new 383 keV feeding transition to 826 keV level along the sequence V was found. 
Justification for its placement is given below. 

\begin{figure}
\begin{center}
\includegraphics[scale=1.4]{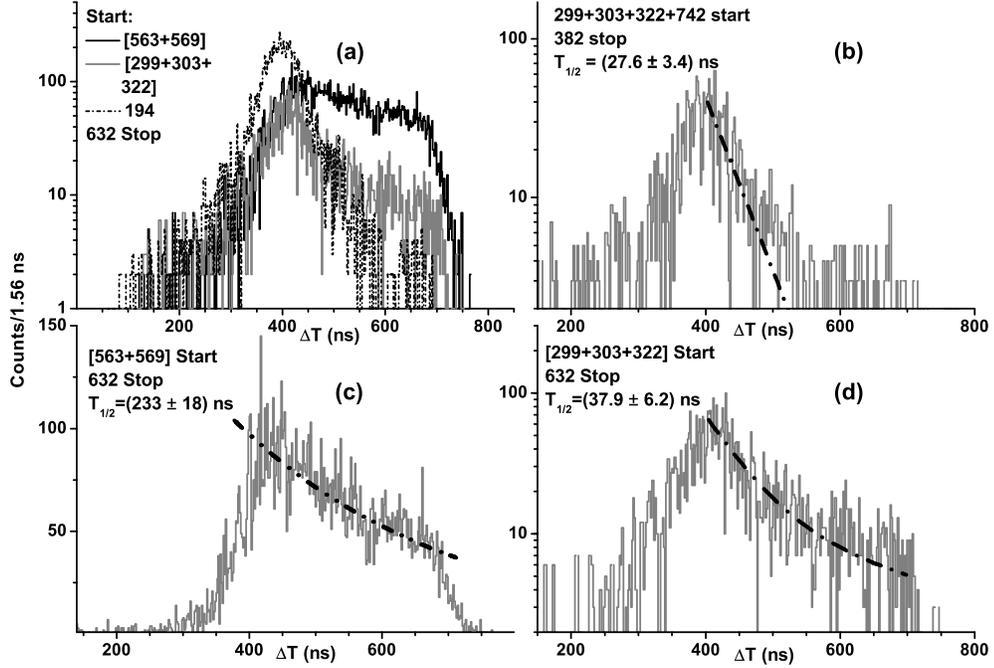} 
\caption{\label{fg7} Different $\Delta T$ spectra used to search for and measure the half lives of 
the isomeric transitions in $^{208}$Fr.}
\end{center}
\end{figure}

One of the major point of controversy is the life time of the 826 keV isomeric level, 
which has been measured by several workers\cite{may,pod,drac}, and the existence of 
other isomers in $^{208}$Fr. A systematic search for isomeric transitions were done 
from our data and the half lives were extracted. 
The $\Delta T$ spectral measurements covered a $\pm 400$~ns TDC range, but an 
useful range of $\sim 500$~ns could be utilized due to delay matching of the detector array. Typical 
$\Delta T$ spectra gated by 632 keV are shown in the Fig.~\ref{fg7}. From the 
194 keV and 632 keV $\gamma$-gated $\gamma \Delta T$ coincidence matrices, we have projected 
the gated $\Delta T$ spectra for the 569 keV, 563 keV, 299 keV, 303 keV, 322 keV and 742 keV gated 
transitions. The $\Delta T$ spectra for the first two gated transitions are found to be similar 
in nature. This clearly establishes the fact that 563 and 569 keV transitions are in sequence 
and above the 826 keV isomeric level. The combined $\Delta T$ spectrum for the $563+569$~keV as start 
and 632 keV as stop are shown in the Fig.~\ref{fg7}(a). However, the $\Delta T$ spectra for the 
299, 303, 322 and 742 keV as start and 194 or 632 keV as stop, though similar in nature among 
themselves, differ significantly from the previous ones (see Fig.~\ref{fg7}(a,d)) in that the 
exponential decay is much faster indicating the existence of another faster isomeric transition 
above the 826 keV level. The gated $\Delta T$ spectra obtained for the gates between the 
pairs of $\gamma$-rays among the 299, 303, 322 and 742 keV, the 563 and 569 keV pair of 
$\gamma$-rays showed their prompt nature. By comparing the prompt and delayed gated spectra, 
taken for the sequence~V transitions mentioned above, a new but weak 383 keV isomeric transition 
was found and it was placed just above the 826 keV level along the sequence~V. 

The half life of the 826 keV isomeric level was extracted by fitting exponential decay function
to the $\Delta T$ spectra, shown in the Fig.~\ref{fg7}(c). The results are given in 
the Table~\ref{tb3}. Half life of $233 \pm 18$~ns was obtained, which is consistent with the 
result $\sim 200$ ns quoted in Ref.~\cite{pod}. However, the result differs from that 
obtained recently by Dracoulis {et al.}\cite{drac}. Half life of the 1209~keV isomeric 
level was extracted from the $\Delta T$ spectra of Fig.~\ref{fg7}(b), in a similar way. The result 
obtained is: $27.6 \pm 3.4$~ns. The same half life was extracted independently from the 
$\Delta T$ spectrum shown in Fig.~\ref{fg7}(d), using two exponential decay fit and the 
measured half life of the 826 keV isomer mentioned above. The results were mutually 
consistent and the average of the two measurements is quoted in Table~\ref{tb3}.    

\section{Interpretation of results}
\label{sc:inter}
The difference in the measured values of the 826 keV isomer half life could not be 
explained. One of the draw back of the present measurement is the restricted range of 
$\Delta T$. This could not be avoided in the experimental set up of the INGA array, which was 
optimized for prompt $\gamma$-spectroscopy. To find out the correctness of the present 
technique, half lives have been extracted for a number of isomeric transitions already 
known in several nuclei produced in our in-beam experiment. These are listed in the 
Table~\ref{tb3}. The isomer half lives spanned from $\sim 150$~ns to $\sim 600$~ns. 
The results are found to be in good agreement with the earlier measurements, within 
the quoted uncertainties.
\begin{table}
\caption{Half lives of isomeric levels of different nuclei produced in the experiment.}
\begin{tabular}{llllll}
\hline\hline
Nucl. & Level & $E_\gamma$ & {\sl ML} & $T_{1/2}$(ns)  & $T_{1/2}$(ns) \\
        & (keV) & (keV) &       & (This expt) & (Earlier) \\ \hline
$^{208}$Fr & 826 & 194.1 &{\sl E1} & 233(18) & 432(11){\protect{\cite{drac}}} \\
 & & & & & $\sim 200$\protect{\cite{pod}} \\
$^{208}$Fr & 1209 & 382.9 &{\sl E2} & 33(7) & \\
$^{208}$Rn & 1828 & 88.7 & {\sl E2} & 590(144) & 509(14)\protect{\cite{trg}} \\
$^{206}$At & 807 & 121.6 &{\sl E1} & 377(44) & 410(80)\protect{\cite{feng}} \\
           &     &       &         &         & 813(21)\protect{\cite{drac}} \\
$^{204}$Po & 1639 & 12.1 & {\sl E2} & 161(4) & 158(2)\protect{\cite{Rahk}} \\
\hline \hline
\end{tabular}
\label{tb3}
\end{table}
The half life for $^{206}$At was also reported in Ref.~\cite{drac} as $813 \pm 21$~ns, which
is almost twice that obtained earlier\cite{feng}. While the probable cause of this difference 
was cited as due to the restricted TDC range $\pm 300$~ns in the previous work, our result does 
not seem to agree with this conclusion. We get results which are consistent with earlier measurements,
though the uncertainties are larger in case of larger half lives (eg. $^{208}$Rn\cite{trg}). The restricted 
range of TDC would result in poor statistics for fitting data, with consequent larger error
bars, as manifested in our results.

The results of DCO ratio measurements are enlisted in the Table~\ref{tb2}. These measurements are 
pivoted on the {\sl E2} assignment of the 632 keV ground state transition, which follows from the 
systematics in the neighbouring nuclei, including Francium isotopes. Based on our DCO results, 
$\Delta J=0$ {\sl E1} character was assigned to the 194 keV isomeric transition. 
The same transition was assigned {\sl E1}($10^- \rightarrow 9^+$) in Ref.~\cite{drac}. However, both the
$J^\pi$ assignments arise from the same $\pi({1h_{9/2}})^{5} \otimes \nu({1i_{13/2}})^{-1}$ configuration, 
with the only difference that the $9^-$ isomeric level would be placed higher than a $10^-$ state, as
evident from the spin multiplets observed in $^{208}$Bi from the neutron pick-up reactions\cite{craw}.
The $11^-$ state, arising out of the same configuration, is assigned to the next 1209 keV level, 
which is also an isomeric level decaying by 383 keV {\sl E2} transition. Based on the measured half life 
and the estimated internal conversion coefficient of 0.0723(11)\cite{kib}, single particle strength 
of 0.0264(12) W.\ u.\ was obtained. For the isomeric transition of the 826 keV $9^-$ level, 
internal conversion coefficient of 0.0947(14) was estimated, and the corresponding single particle 
strength was obtained as $1.06(8)\times 10^{-7}$ W.\ u. Similar results for the {\sl E1}, 
{\sl M1}, {\sl E2} and {\sl M2} isomeric transitions are obtained in  
trans-Lead nuclei(eg.\ $^{208}$Rn\cite{trg}) in this mass region. 

In the level scheme shown in Fig.~\ref{fg4}, the series of levels from 632 keV ($7^+$) to 
1983 keV ($15^+$) along sequence I arise from the $\pi (1h_{9/2})^5 \otimes \nu (2f_{5/2})^{-1}$ with 
proton excitations of higher seniority leading to the generation of angular momenta. This 
is clear from the connecting sequence of {\sl E2} transitions. The maximum angular momentum generated 
from proton excitation in this case is ${25/2}$, which leads to maximum $J^\pi = 15^+$ for the 
given configuration. The 123 and 156 keV transitions above the 1983 keV level could be of 
{\sl E2} nature, though our data is inadequate to make a definite conclusion.

The sequence II transitions, observed for the first time, are weak in intensity but they are 
fitted into the level scheme from the matched sequence of $\gamma$-ray energies, coincidence 
conditions and intensities. 
However, the spin assignments of only the 2603 keV level could be done from the DCO ratio for 
the 645 keV transition by gating on 563 keV {\sl M1} transition. This could be a stretched 
{\sl E2} transition or a $\Delta J=0$ {\sl M1} transition, but the latter assignment was adopted
to fix the spins of the level sequence. The 604 keV, 499 keV and 150 keV transitions appear to 
be {\sl M1} though it cannot be confirmed from our data.
Sequence III transitions extend to the highest excitation energy of $^{208}$Fr in this experiment.
It starts from the 1395 keV $11^-$ level which is connected by 569 keV {\sl E2} transition to the 
826 keV $9^-$ level. This sequence is likely to arise from the $\pi ({1h_{9/2}})^4 ({2f_{7/2}}) \otimes 
\nu ({1i_{13/2}})^{-1}$ configuration which leads to the highest spin of $22^{-}$, and are connected 
by a series of {\sl M1} transitions. Though we could extend up to $20^-$ level, the absence of $19^-$
level is possibly because of the fact that it is pushed down by the residual proton particle neutron hole 
repulsive interaction, from where gamma transitions are hindered. A detailed shell model calculation
will be needed to make a definite conclusion in this regard.

The low lying transitions of the sequences V and VI, extending up to 2763 keV $J=19$ level can be formed 
by $\pi ({1h_{9/2}})^5 \otimes \nu ({1i_{13/2}})^{-1}$ configuration through proton excitation. 
However, levels above $19^-$ along sequence VI, built on proton excitation to the maximum proton spin 
of $25/2$, are due to neutron hole excitation arising from $\nu ({2f_{5/2}}^{-2} 1i_{13/2}^{-1})$, 
leading up to the maximum $J^\pi = 23^-$ level observed. The series of {\sl E2} transitions above 
the 2262 keV $15^-$ level along the sequence V can be between levels generated by 
$\pi ({1h_{9/2}}^4 {1i_{13/2}}) \otimes \nu ({2f_{5/2}})^{-1}$, which leads to a maximum 
$J^\pi=21^-$. The levels above it are connected by {\sl M1} transitions which are probably 
generated by a different configuration. A better statistics and / or higher resolving power 
of the array will be needed to extend the level scheme further.      

\section{Conclusion}
\label{sc:concl}

The level scheme of $^{208}$Fr was modified over the existing level scheme, and extended  up to 
$\sim 23 \hbar$ and $\sim 5.5$ MeV excitation energy using a high resolving power Clover detector 
array. A number of new $\gamma$-transitions were observed and their DCO ratios were measured. 
Based on search for isomeric transitions from the data using the tagged $\gamma \gamma$ time 
difference technique, the half lives of several isomeric transitions in $^{208}$Fr and in a few 
neighbouring nuclei produced as ER in the experiment were measured. The results agree reasonably 
well with the previously known half lives. The half life of the known 194 keV isomeric 
transition in $^{208}$Fr was found to differ from the previously reported value. A new isomeric 
{\sl E2} transition was obtained and its half life was measured. The Weisskopf estimate of 
the single particle strength of the associated isomeric levels reveal similarity with the 
previous estimates in the neighbouring nuclei. From the shell model based interpretation of 
the level scheme, it is clear that the majority of the excited states are caused by $1h_{9/2}$ 
proton excitations, and neutron hole excitations predominantly in $2f_{5/2}$ and $1i_{13/2}$ 
shells. The importance of $1i_{13/2}$ neutron hole in generating the isomeric levels is 
clear from the present data. It may be noted that though we have observed only two isomer 
levels, there can be a few more such levels which could not be observed due to limted 
statistics in our data. A pulsed beam based experiment, coupled with such a 
high resolving power array will be needed to extend the study further. 

\section{Acknowledgement}
\label{sc:ackn}
We are grateful to all the colleagues of the INGA collaboration for their help 
during the experiment. Smooth running of the 16UD Pelletron accelerator and 
the INGA detector array at the IUAC, New Delhi by the staff therein are 
gratefully acknowledged.


\end{document}